# Array Configuration-Agnostic Personal Voice Activity Detection Based on Spatial Coherence

Yicheng Hsu[1]
Department of Power Mechanical Engineering, National Tsing Hua University
101, Section 2, Kuang-Fu Road, Hsinchu, Taiwan 30013

Mingsian R. Bai[2]
Department of Power Mechanical Engineering and Electrical Engineering, National Tsing Hua University
101, Section 2, Kuang-Fu Road, Hsinchu, Taiwan 30013

**ABSTRACT**

*Personal voice activity detection (PVAD) has received increased attention due to the growing popularity of personal mobile devices and smart speakers. PVAD is often an integral element to speech enhancement and recognition for these applications in which lightweight signal processing is only enabled for the target user. However, in real-world scenarios, the detection performance may degrade because of competing speakers, background noise, and reverberation. To address this problem, we proposed to use equivalent rectangular bandwidth (ERB)-scaled spatial coherence as the input feature to train an array configuration-agnostic PVAD (ARCA-PVAD) network. Whereas the network model requires only 112k parameters, it exhibits excellent detection performance and robustness in adverse acoustic conditions. Notably, the proposed ARCA-PVAD system is scalable to array configurations (geometry, number of microphones, and spacing). Experimental results have demonstrated the superior performance achieved by the proposed ARCA-PVAD system over a baseline in terms of the area under receiver operating characteristic curve (AUC) and equal error rate (EER).*

## 1. INTRODUCTION

Standard voice activity detection (VAD) is often an important front-end of modern speech processing such as automatic speech recognition and speaker identification. Standard VAD systems use classifiers on the acoustic features to obtain speech/non-speech predictions for each audio frame. Conventional approaches typically make use of energy-related features and statistical model-based metrics, including signal-to-noise ratio (SNR) and likelihood ratio [1], or speech-specific features such as linear prediction coefficients [2], zero-crossing rate [3], and cepstral features [4]. Whereas these VAD approaches generally perform satisfactorily for high SNR scenarios, the detection performance can degrade significantly when SNR is low. Recently, deep-learning-based VAD techniques [5, 6, 7, 8, 9, 10] have received much attention, but their performance can be impacted in the presence of speech-like interferences. It is then desirable to develop a high performance but robust VAD for application scenarios with adverse acoustical conditions. Furthermore, a multichannel VAD can be more advantageous than the single-channel counterparts by leveraging the spatial features such as interchannel time, level, and phase differences [11, 12] provided by the microphone array.

The idea of personal VAD (PVAD) was reported in [13]. In some previous works [13, 14, 15], only the voice activity of a target speaker is detected using the auxiliary information. Speaker embedding such as the i-vector [16] or the d-vector [17] can be obtained using pre-enrolled utterances of the target speaker. The need for pre-enrollment can be waived if special schemes are employed for training the PVAD [18]. Although these approaches are effective in extracting the target speech, the detection performance often degrades due to speech-like interferences such as competing speakers or TV dialogues.

---

[1] shane.ychsu@gmail.com
[2] msbai@pme.nthu.edu.tw

In this study, we propose an array configuration-agnostic PVAD (ARCA-PVAD) system by exploiting speaker embeddings derived from user enrollment and equivalent rectangular bandwidth (ERB)-scaled long short-term spatial coherence (LSTSC) proposed in [19, 20]. The ERB-scaled spatial coherence is computed as the spatial features associated with speaker activities for each time–frequency bin. Furthermore, the proposed system generalizes to unseen accommodate array geometries and channel counts than those used in the training phase. To assess the robustness of the proposed approach, unseen array geometries and number of microphones are evaluated in the testing phase via a series of experiments. The area under the receiver operating characteristic curve (AUC) and the equal error rate (EER) are adopted as the performance metrics. A personal VAD reported in [13] as well as a single-microphone system are employed as the baselines to benchmark the proposed ARCA-PVAD model.

## 2. Proposed ARCA-PVAD system

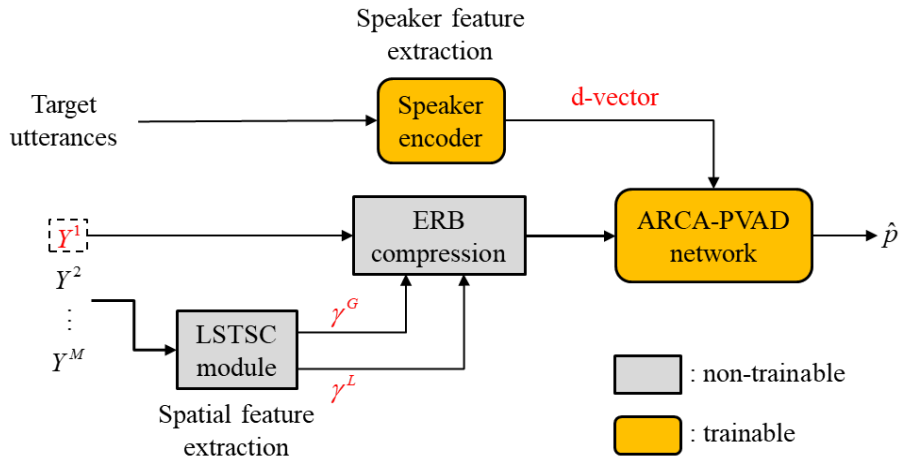

Figure 1: The proposed ARCA-PVAD system.

### 2.1. Problem formulation and signal model

In this section, we present an ARCA-PVAD system that detects the target speaker activity from multichannel signals using a target speaker's embedding. Consider a reverberant room with one static interference source, the target speaker, and a non-target speaker. The signals emitted by the sources are captured by a microphone array containing $M$ elements. The array signal model can be expressed in the short-time Fourier transform (STFT) domain, with $l$ and $f$ denoting the time frame index and the frequency bin index, respectively. The signal captured by the $m$th microphone can be written as

$$Y^m(l,f) = \sum_{j=1}^{2} A_j^m(f) S_j(l,f) + A_n^m(f) N(l,f) + V^m(l,f), \qquad (1)$$

where $A_j^m(f)$ denotes the acoustic transfer function between the $m$th microphone and the $j$th speaker, $A_n^m(f)$ denotes the ATF between the $m$th microphone and the interferer, $S_1(l,f)$, $S_2(l,f)$, and $N(l,f)$ denote the STFTs of the target signal, the non-target speaker, and the interferer, and $V^m(l,f)$ denotes the additive sensor noise measured by the $m$th microphone.

### 2.2. Overall system architecture

To tackle the personal VAD problem stated above, we proposed an ARCA-PVAD system to detect the activity of the target speaker from the noisy mixture signals. Figure 1 demonstrates the overall system architecture, which consists of three modules: the spatial feature extraction module (Section

2.3) based on LSTSC, the speaker feature extraction module (Section 2.4) based on the pre-enrolled utterances of the target speaker, and the ARCA-PVAD network based on the auxiliary information provided by the two previous modules. All modules will be detailed next.

**2.3. The LSTSC feature**

For each TF bin, a "whitened" RTF vector $\mathbf{r}(l,f) \in \mathbb{R}^{M-1}$, with the first entry deleted, can be used as a short-term spatial feature:

$$\mathbf{r}(l,f) = \left[ \frac{R^2(l,f)}{|R^2(l,f)|}, \ldots, \frac{R^M(l,f)}{|R^M(l,f)|} \right]^T, \quad (2)$$

where $R^m(l,f)$ denotes the short-term RTF between the $m$th microphone and the reference micropone 1. For spatially stationary interferer, the following long-term RTF is computed via recursive averaging:

$$\bar{r}^m(l,f) = \lambda \bar{r}^m(l-1,f) + (1-\lambda) r^m(l,f), \quad m=2,\ldots,M, \quad (3)$$

where $0 < \lambda < 1$ is the forgetting factor that regulates the averaging process. The long-term RTF vector $\bar{\mathbf{r}}(l,f)$ is also whitened to serve as a long-term spatial feature vector:

$$\bar{\mathbf{r}}(l,f) = \left[ \frac{\bar{r}^2(l,f)}{|\bar{r}^2(l,f)|}, \ldots, \frac{\bar{r}^M(l,f)}{|\bar{r}^M(l,f)|} \right]^T. \quad (4)$$

The LSTSC feature, $\gamma(l,f)$, between the short-term whitened feature vector $\mathbf{r}(l,f)$ and the long-term whitened feature vector $\bar{\mathbf{r}}(l,f)$ can be calculated as follows:

$$\begin{aligned}\gamma(l,f) &= \frac{\operatorname{Re}\{\mathbf{r}^H(l,f)\bar{\mathbf{r}}(l,f)\}}{\sqrt{\mathbf{r}^H(l,f)\mathbf{r}(l,f)}\sqrt{\bar{\mathbf{r}}^H(l,f)\bar{\mathbf{r}}(l,f)}} \\ &= \frac{1}{M-1}\operatorname{Re}\{\mathbf{r}^H(l,f)\bar{\mathbf{r}}(l,f)\}\end{aligned} \quad (5)$$

where $\operatorname{Re}\{\cdot\}$ denotes the real-part operator and the ($M$-1) factor in the denominator results from the fact that the feature vectors have been whitened. In addition, we define the LSTSC with a large forgetting factor ($\lambda = 0.99$) as the global LSTSC ($\gamma^G$), whereas the LSTSC with a small forgetting factor ($\lambda = 0.01$) are defined as the local LSTSC ($\gamma^L$)

Considerable computational savings can be achieved given the equivalent rectangular bandwidth (ERB) [21]. The spectrum of a signal and the LSTSC feature can be represented in a few $B$ bands:

$$Y_{ERB}(l,b) = \sum_{f \in \{f_{b1},\ldots,f_{bF_b}\}} w_b(f) |Y^1(l,f)|^2, \quad b \in \{0,1,\ldots,B\} \quad (6)$$

$$\gamma_{ERB}(l,b) = \frac{1}{\pi_b} \sum_{f \in \{f_{b1},\ldots,f_{bF_b}\}} w_b(f) \gamma(l,f), \quad b \in \{0,1,\ldots,B\}, \quad (7)$$

where $w_b(f)$ and $F_b$ are the weight and the number of the frequency bins for band $b$, respectively. The weight normalization factor

$$\pi_b = \sum_f^{F_b} w_b(f), \tag{8}$$

with the ERB transformation, the dimension of the LSTSC feature can be reduced to $B$ bands ($B = 32$ in this paper).

### 2.3. Speaker encoder

The speaker encoder generates speaker embeddings from the pre-enrolled utterances of the target speaker. In this study, we used the d-vector [17], widely used in personal VAD, speech translation, speaker diarization, etc. The speaker encoder consists of a three-layer long short-term memory (LSTM) network and is trained using a generalized end-to-end loss function suggested in a previous study [17]. The speaker encoder was trained using the VoxCeleb2 data set [22]. The model yields embeddings in terms of sliding windows. The resulting aggregated embedding, known as the d-vector, encodes the voice characteristics of the target speaker.

### 2.3. ARCA-PVAD network

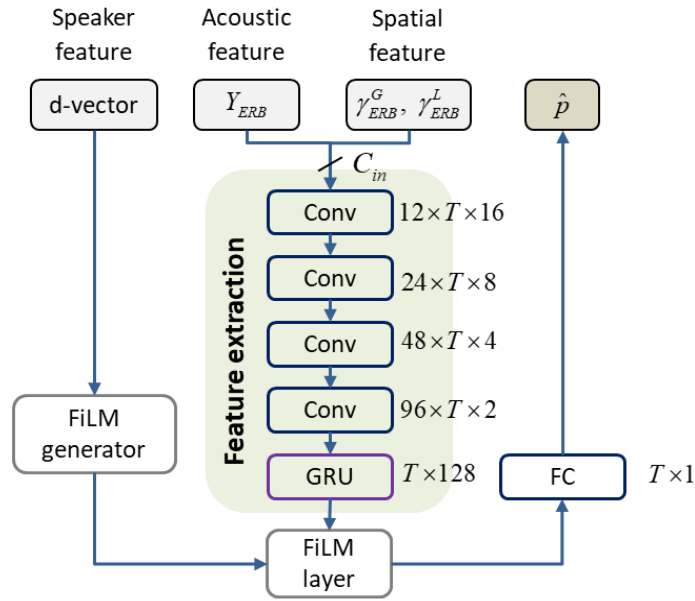

Figure 2: The ARCA-PVAD network.

Building on the success of the convolutional recurrent network (CRN) structure for efficient VAD [23], we use a similar design for the PVAD task. Figure 2 illustrates the proposed network architecture. The ARCA-PVAD model incorporates three input features: the ERB-scaled spectral feature of reference, the ERB-scaled LSTSC features, and the d-vector obtained by the speaker decoder. The speaker information is detected by the convolutional neural networks (CNNs) and the gated recurrent unit (GRU) layer. The extracted speaker feature is then modulated via a feature-wise affine transformation (FiLM) layer [24]. Finally, the output of the FiLM layer is passed to the final fully-connected classifier. As illustrated in Fig. 2, the four CNN layers with a 12-24-48-96 filter. The input channel $C_{in}$ is set to 3 in this study. The convolutional blocks consist of a separable convolution, followed by batch normalization and ReLU activation. The convolution kernel and step size are set to (3,2) and (2,1). The output layer uses a sigmoid to obtain constrained values.

## 3. Experimental study

### 3.1. Settings of experiment

Clean utterances containing utterances from 921 and 40 speakers were selected from the train-clean-360 and test-clean subsets of the LibriSpeech corpus [25] for training and testing. The VoxConverse dataset [26], which consists of 74 h of human conversation clips from YouTube[@], was used to simulate the speech-like TV interferences. The experiment was performed at a sample rate of 16 kHz.

In the training phase, 50,000 and 5,000 samples were used in training and validation. Noisy signals edited in seven-second clips were prepared by mixing the target speech signal and speech-like TV noise with signal-to-interference ratio (SIR) = -10, -5, 0, 5, 10, and 15 dB. The target speech signal and the non-target speech signal were set to be equal in power. Sensor noise was added with signal-to-noise ratio (SNR) = 30 dB. In addition, the target and non-target speaker signals were not overlapped in time. As depicted in Fig. 3, the target speaker, speech-like TV interference, and non-target speaker were randomly placed in the frontal plane in the ring with radius = 1 and 2 m. The target speaker and non-target speaker were set to be closer to the array center than the TV interferer. In addition, any two sources are at least 15° apart from each other. Three four-element linear arrays shown in Fig. 4 were used for training and validation. Reverberant microphone signals were simulated by convolving the dry signals with Room impulse responses (RIRs) by using the Multi-Channel Impulse Responses Database [27], which recorded with the reverberation time (T60) = 0.16, 0.36, and 0.61 s at Bar-Ilan University.

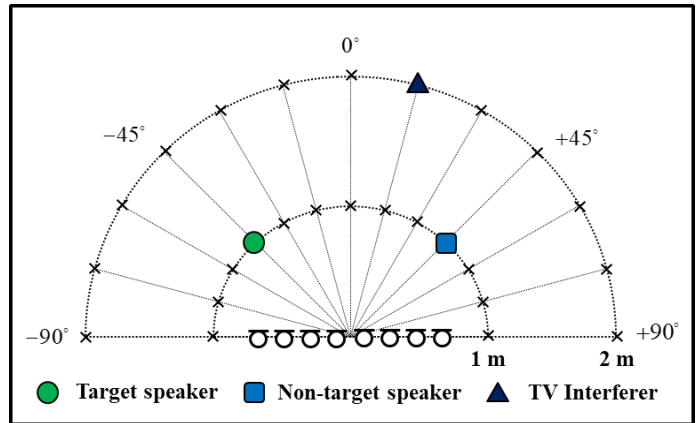

Figure 3: Illustration of the experimental settings for training.

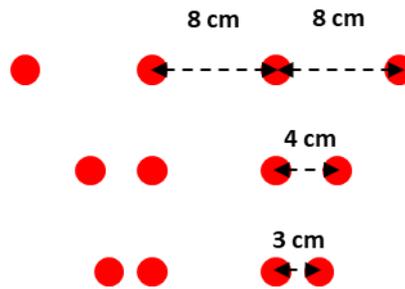

Figure 4: Array geometries used as the training set.

In the test set, we generated a 1,600-sample dataset to validate the robustness of the proposed ARCA-PVAD system when applied to unseen RIRs, unseen array geometries, and the unknown number of microphones. We created the test set based on the Tampere University Rotated Circular Array Dataset [28] recorded using a seven-element microphone array, with one microphone at the center and the others on a circle with a radius of 4 cm. The RIRs are measured from −90° to 90° in 5° intervals at a distance of 0.7, 1.4, and 2.1 m from the array center. In addition, any two sources are at least 15° apart from each other. In addition, eight array configurations were employed to evaluate the robustness of the proposed ARCA-PVAD system. An illustration of these array configurations is depicted in Fig. 5. The reference microphone is marked green.

For the training of the baseline and the proposed models, the signal frame was selected to be 25 ms long with a 10 ms step. A 512-point FFT was used. The optimizer was Adam, with a learning rate of 0.001 and a gradient norm clipping of 3. The learning rate was halved if the loss of the validation set did not improve for three consecutive epochs. We trained the baseline model using only the ERB-scaled acoustic feature. In addition, we also consider that the proposed ARCA-PVAD system should be able to achieve satisfactory performance while the spatial information is unavailable. Inspired by Ding et al. [14], who proposed the training paradigm for both enrollment and enrollment-less. During the training epoch, we sampled a subset of training samples with probability $p^* = 0.1$. For the sampled subset, we set the LSTSC features to 1. In this way, the proposed ARCA-PVAD system can also handle single-channel cases.

For all datasets, the frame-level ground-truth personal VAD labels for clean utterances of the target speaker were obtained by considering the threshold (in decibels). When the energy of the target clean utterance was above the reference threshold, the output was 1; for energies below the threshold, the output was 0.

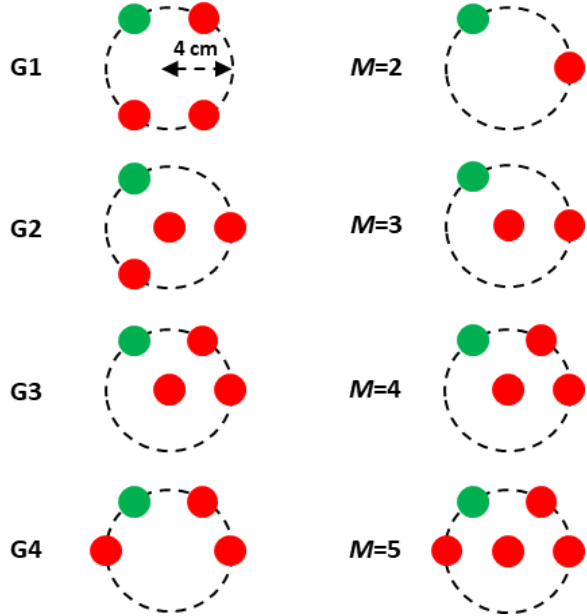

Figure 5: Different array configurations for the test set.

### 3.2. Results

To evaluate the performance of the proposed ARCA-PVAD system, we computed two metrics, namely AUC and EER, which equate to the point at which the false positive rate (FPR) and false negative rate (FNR) cross each other. First, we evaluate the performance of the personal VAD with a variety of array geometries. Figure 6 shows the AUC and EER performance for different array geometries. The results demonstrate that the proposed ARCA-PSE system has higher AUC and lower EER than the baseline by exploiting the LSTSC features, especially at lower SIR cases. Furthermore, even if the array geometries were not included in the training set, the ARCA-PSE system still performs considerably. Second, we examine the effect of the sensor count on personal VAD. As shown in Fig. 7, the personal VAD performance is significantly improved in both AUC and EER with an increase in the number of microphones, even if we only trained using a four-element array. Furthermore, with a simple training paradigm, the proposed ARCA-PVAD is also able to handle single-channel cases. In the single-channel case, the ARCA-PVAD system has comparable performance to the baseline. Without requiring changes to the model architecture for each array configuration, a single model trained with the proposed LSTSC function can be shared among multiple arrays with different shapes and numbers of microphones.

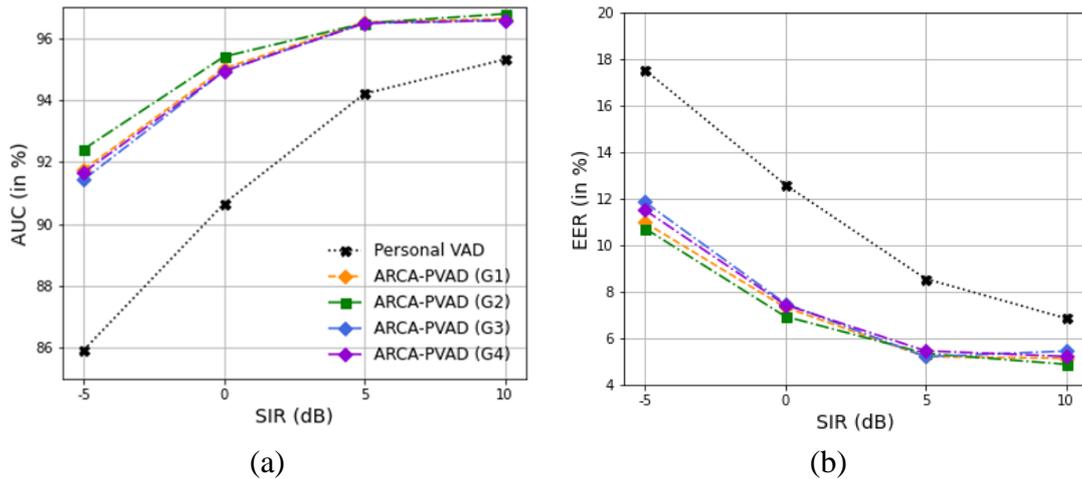

Figure 6: Comparisons of PVAD performance in terms of (a) AUC and (b) EER for different array geometries.

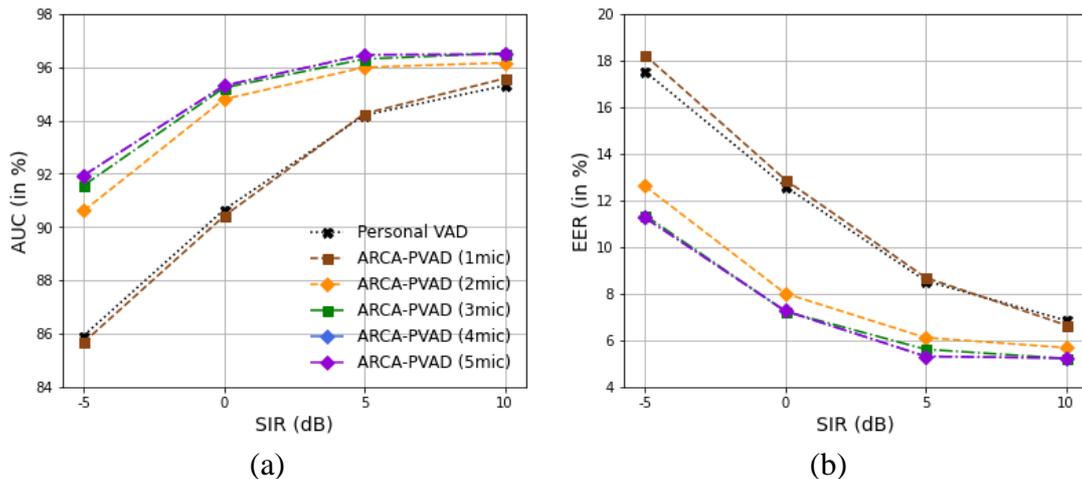

Figure 6: Comparisons of PVAD performance in terms of (a) AUC and (b) EER for different numbers of microphones.

## 4. Conclusions

In this study, we have proposed a lightweight network model ARCA-PVAD on the basis of acoustic features, the ERB-scaled spatial feature, and the target speaker embedding for detection of voice activities in adverse conditions. Compared with the single-channel model that only considers acoustic features and target speaker embedding as inputs, the proposed model that exploits additionally the spatial features shows improved performance and robustness for unseen RIRs, array geometries, and numbers of microphones, which is highly desirable for real-world applications. With the proposed training paradigm, the ARCA-PVAD system can also behave as a single-channel personal VAD when spatial information is unavailable.

## Acknowledgements

This work was supported by the National Science and Technology Council (NSTC) in Taiwan, under the project number 110-22221-E-007-0227-MY3.